# Is it possible to consider a thin hologram as a slide?

Anatoly M. Smolovich

Kotel'nikov Institute of Radio Engineering and Electronics (IRE) of the Russian Academy of Sciences, Mokhovaya 11-7, Moscow 125009, Russia
E-mail: asmolovich@petersmol.ru

**ABSTRACT**

A thin phase hologram with sinusoidal modulation of the refractive index is considered. The applicability of the approach is discussed, in which it is assumed that light passes through a hologram, as through a slide, which leads to a sinusoidal modulation of the wave phase. A consequence of this consideration are expressions for the diffraction orders amplitudes, which coincide with the expressions obtained in the Raman-Nath theory of light scattering by ultrasonic waves. The paper compares expressions for the amplitude of the first order of diffraction, obtained in the framework of several theoretical approaches, including perturbation theory in a rigorous formulation of the boundary value problem. As the thickness of the hologram tends to zero, the expression obtained in this way coincides with the expression obtained in the two-wave approximation of the coupled-wave theory. Both expressions satisfy the electrodynamic reciprocity theorem. Under the same conditions, the expression obtained as a result of the slide-like consideration does not satisfy the reciprocity theorem and differs from the indicated expressions by a factor equal to the ratio of the cosines of the diffracted and incident waves. It is concluded that considering a thin hologram as a slide can lead to noticeable errors.

**Keywords:** thin hologram, wave equation, Raman-Nath diffraction, perturbation theory, coupled-wave theory

## 1. Introduction

We are interested in approaches that make it possible to obtain simple approximate analytical expressions for the amplitudes of the diffraction orders of a thin hologram with a sinusoidal modulation of the refractive index. At first glance, it seems that the passage of a light wave through a hologram whose thickness tends to zero can be represented as the multiplication of the wave amplitude by some two-dimensional complex transmission function. In other words, it seems that the process is similar to the passage of a beam of light from a slide projector through a slide. With this consideration, the phase of the transmitted wave acquires a sinusoidal modulation [1]. The field in the half-space behind the grating can then be represented as a superposition of plane waves (diffraction orders) with amplitudes proportional to Bessel functions of the corresponding order [2-6]. The expressions obtained in this manner coincide with the well-known Raman-Nath expressions for the diffraction order amplitudes [7]. The use of the Raman-Nath expressions for thin holograms raises the following questions: a) the Raman-Nath theory derived for the case when the grating period significantly exceeds the wavelength of light, and in holography, these are usually values of the same order; b) the Raman-Nath theory was constructed for volume periodic structures, and for thin holograms, it is assumed that the hologram thickness is negligibly small. In addition, it turned out that the use of the Raman-Nath expressions for the amplitudes of waves diffracted by each layer in the thin-grating decomposition method [2] can lead to an energy imbalance [8].

To understand the above issues, in this study we consider the problem of light diffraction by a hologram of finite thickness with a sinusoidal modulation of the refractive index. A rigorous statement of the boundary-value problem is used, taking into account the transmitted and reflected diffraction orders [9, 10]. The wave equation is solved using perturbation theory [11-14]. We compare the expression for the first diffraction order obtained in the first approximation of the perturbation theory with the corresponding expression obtained in simplified consideration, as the hologram thickness ($d$) tends to zero. To achieve this, both expressions are expanded to a series in $d$ and limited to a linear term. Both expressions are analyzed from the perspective of the fulfillment of the electrodynamic reciprocity theorem [15, 12]. For a particular case, the expressions are compared with the results of the two-wave approximation of coupled-wave theory [16]. Their use in the thin-grating decomposition method [2] is also discussed.

## 2. Simplified consideration of the passage of a plane wave through a thin hologram

Let a thin hologram be limited by the planes $z = 0$ and $z = d$, the refractive index of the hologram has a sinusoidal modulation $n = n_0 + n_1 \cos(\frac{2\pi x}{\Lambda})$, where $\Lambda$ is the grating period (Fig. 1). Let a plane wave with a wave vector **k** lying in the $xz$ plane falls on the hologram from the region $z<0$ (Fig. 1), where $k = k_0 n_0 = \frac{2\pi n_0}{\lambda}$, and $\lambda$ is the wavelength of light. It is usually assumed that a wave passing through a thin phase sinusoidal hologram acquires sinusoidal phase modulation; that is, up to a constant factor, the field in the $z=d$ plane is equal to (Section 8.5 in [1]):

$$\exp[i\xi \cos(\frac{2\pi x}{\Lambda})], \qquad (1)$$

where $\xi$ is a constant defined below. Next, we consider the fact that the function

$$\exp[ixu + iz(k^2 - u^2)^{1/2}], \qquad (2)$$

where $u$ is arbitrary, is the exact solution of the wave equation [17] for $z>d$ [18]. Owing to the linearity of the wave equation, its solution is also the sum

$$\sum_{m=-\infty}^{\infty} i^m J_m(\xi) \exp\left\{ix(k_x + \frac{2\pi m}{\Lambda}) + iz[k^2 - (k_x + \frac{2\pi m}{\Lambda})^2]^{1/2}\right\}, \qquad (3)$$

where $J_m$ is an $m$th-order Bessel function. Expression (3) exactly satisfies the unperturbed wave equation in the region $z>d$, and coincides with (1) in the plane $z = d$ (Section 7.2.4 [19]). To determine the specific values of the amplitudes of the diffraction orders, $\xi$ must be related to hologram parameters. It is generally believed [2-6] that

$$\xi = \frac{k_0 n_1 d}{\cos\theta}, \qquad (4)$$

where θ is the angle between the vector $k$ and the normal to the hologram. This expression is obtained if we assume that the incident wave passes through a thin hologram as through a slide, i.e. a plane wave propagates inside the hologram according to the laws of geometric optics, which gives a complete phase shift $\frac{k_0 d}{\cos\theta}\left[n_0 + n_1 \cos(\frac{2\pi x}{\Lambda})\right]$. Then from (3, 4) we can obtain the following expression for the amplitude of the *m*-th diffraction order:

$$E_m = (i)^m J_m\left(\frac{2\pi d n_1}{\lambda \cos\theta}\right). \qquad (5)$$

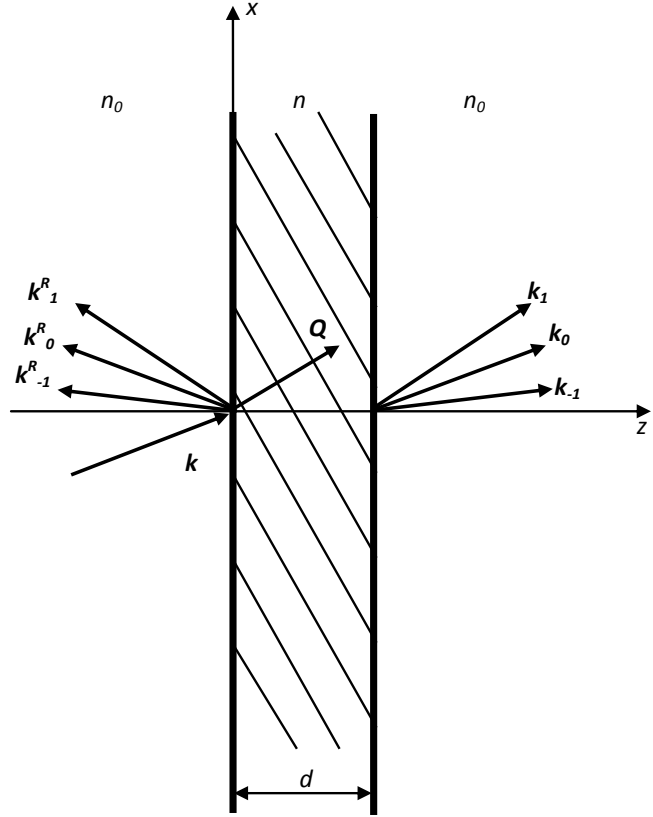

Fig. 1. The geometry of a phase grating with an incident plane wave and diffracted waves. The wave vectors are denoted as follows: $k$ is the wave vector of the incident wave, $k_{-1}, k_0, k_1$ are the wave vectors of the transmitted diffraction orders; $k^R_{-1}, k^R_0, k^R_1$ are the wave vectors of the reflected diffraction orders; $Q$ is the grating vector.

Expression (21) coincides with the Raman-Nath expression for the amplitudes of diffraction orders [7]. The use of Raman-Nath expressions here raises the following questions: a) Raman-Nath expressions obtained for the case when the grating period significantly exceeds the wavelength of light, and in holography these are usually values of the same order; b) the Raman-Nath theory is constructed for bulk periodic structures, and the thin hologram thickness is assumed to be negligible. To answer these questions, the next section uses a rigorous formulation of the boundary value problem, which is solved using perturbation theory.

3. **Solution of the wave equation in the first approximation of perturbation theory**

Consider a flat phase hologram of thickness $d$ with sinusoidal modulation of refractive index $n$ and grating vector $\boldsymbol{Q}$ in the $xz$ plane (Fig. 1). We assume that the $x$-axis lies in the hologram plane, and that the $z$-axis coincides with the normal of the hologram. In this case, in the region $0 \leq z \leq d$, the refractive index is

$$n = n_0 + n_1 \cos(\boldsymbol{Qr}), \tag{6}$$

where $\boldsymbol{r}$ denotes the radius vector. Let a linearly polarized plane wave with a wave vector $\boldsymbol{k}$ lying in the $xz$ plane and an electric vector having only the $y$-component falls on the hologram from the region $z<0$. In this case, the resulting electric field also has only the $y$-component $E_y$, and the problem becomes scalar. $E_y$ must satisfy the following wave equation [17].

$$\nabla^2 E_y + n^2 k_0^2 E_y = 0. \tag{7}$$

In (7), time factor $\exp(-i\omega t)$ is omitted. The local refractive index of medium $n$ in (7) is described by (6) within the hologram ($0 \leq z \leq d$) and $n = n_0$ outside ($z<0$, $z>d$). The continuity conditions of the tangential components of the vectors of the electric and magnetic fields were used as boundary conditions at $z=0$ and $z = d$. It can be obtained from Maxwell's equations [17] that the tangential component of the magnetic field, $H_x$, is proportional to $\dfrac{\partial E_y}{\partial z}$.

Therefore, in our case, the continuity conditions of $E_y$ and $\dfrac{\partial E_y}{\partial z}$ are used as the boundary conditions. We will solve this problem using perturbation theory [12], which represents the electric field of the wave in the region $0 \leq z \leq d$ as follows[1]:

$$E_y = \exp(i\boldsymbol{kr})[E^{(0)} + E^{(1)} + E^{(2)} + \ldots]. \tag{8}$$

According to the usual practice of perturbation theory, it is assumed that $E^{(0)} \gg \quad \gg \quad \gg \quad$. In this case, $E^{(0)}\exp(i\boldsymbol{kr})$ must satisfy the unperturbed ($n=n_0$) equation (7), and $E^{(0)}$ is a constant assumed to be equal to unity. The equation for the first perturbation theory approximation follows from (6-8):

$$\frac{\partial^2 (E^{(1)})}{\partial x^2} + \frac{\partial^2 (E^{(1)})}{\partial z^2} + 2i\left[k_x \frac{\partial (E^{(1)})}{\partial x} + k_z \frac{\partial (E^{(1)})}{\partial z}\right] = -k_0^2 n_0 n_1 \left[\exp(i\boldsymbol{Qr}) + \exp(-i\boldsymbol{Qr})\right]. \tag{9}$$

We will look for $E^{(1)}(x,z)$ in the form:

$$E^{(1)}(x,z) = E_1(z) \exp(iQ_x x) + E_{-1}(z) \exp(-iQ_x x). \tag{10}$$

From (9) and (10), we obtain an ordinary differential equation with constant coefficients for the wave amplitude $E_1(z)$ corresponding to the (+1)th diffraction order as follows:

$$\frac{d^2 E_1}{dz^2} + 2ik_z \frac{dE_1}{dz} - (Q_x^2 + 2k_x Q_x) E_1 = -k_0^2 n_0 n_1 \exp(iQ_z z), \tag{11}$$

---

[1] We use expression (8), which is somewhat different from the similar expression in [12].

A general solution of (11) is:

$$E_1 = E^{(0)} G \exp(iQ_z z) + C_1 \exp[iz(-k_z + k_{1z})] + C_2 \exp[iz(-k_z - k_{1z})], \quad (12)$$

where

$$G = \frac{k_0^2 n_0 n_1}{Q^2 + 2(\mathbf{k}\mathbf{Q})}. \quad (13)$$

Here, the components of the wave vectors of the diffraction orders are determined by the expressions

$$k_{1x} = k_x + Q_x,$$

$$k_{1z} = [k^2 - (k_x + Q_x)^2]^{1/2}. \quad (14)$$

Constants $C_1$ and $C_2$ in (12) must be determined from the boundary conditions. On the plane $z = d$, the field $E_1(z)$ must be "matched" with the field of a plane wave in a homogeneous medium:

$$E_1^T \exp[i(k_{1x} x + k_{1z} z)], \quad (15)$$

corresponding to the 1st transmitted diffraction order. On the plane $z=0$, the field $E_1(z)$ must be "matched" with the field of the plane wave:

$$E_1^R \exp[i(k_{1x} x - k_{1z} z)], \quad (16)$$

corresponding to the 1st reflected diffraction order. "Matching" here means equating functions and their first derivatives with respect to z. This yields the following system of four equations.

$$E_1^R = GE^{(0)} + C_1 + C_2, \quad (17)$$

$$-ik_{1z} E_1^R = i(k_z + Q_z) GE^{(0)} + ik_{1z} C_1 - ik_{1z} C_2, \quad (18)$$

$$E_1^T \exp(ik_{1z} d) = GE^{(0)} \exp[i(k_z + Q_z)d] + C_1 \exp(ik_{1z} d) + C_2 \exp(-ik_{1z} d), \quad (19)$$

$$ik_{1z} E_1^T \exp(ik_{1z} d) = i(k_z + Q_z) GE^{(0)} \exp[i(k_z + Q_z)d] + ik_{1z} C_1 \exp(ik_{1z} d) - ik_{1z} C_2 \exp(-ik_{1z} d), \quad (20)$$

From the system of equations (17-20) constants $C_1$, $C_2$, $E_1^T$, and $E_1^R$ are found. In particular, the amplitude of the 1st transmitting diffraction order is equal to

$$E_1^T = \frac{G(k_{1z} + k_z + Q_z)\{\exp[i(-k_{1z} + k_z + Q_z)d)] - 1\}}{2k_{1z}}. \quad (21)$$

### 4. Comparison of expressions for the amplitude of the first diffraction order obtained using several theoretical approaches

Let us compare, at small $d$, the amplitudes of the 1st diffraction order, calculated by expression (5) at $m=1$, and by expression (21), obtained on the basis of perturbation theory. To do this, we expand both expressions to a series in *d*, restricting ourselves to a linear term, denoting it as $^{RN}E_1^T$ and $^{PT}E_1^T$, respectively.

$$^{RN}E_1^T = \frac{ik_0 n_1 d}{2\cos\theta}, \tag{22}$$

$$^{PT}E_1^T = \frac{ik_0 n_1 d}{2\cos\theta_1}, \tag{23}$$

where $\cos\theta_1$ is the cosine of the slope of the wave vector of the 1st diffraction order to the *z*-axis. The denominator of (22) contains $\cos\theta$, and the denominator (23) contains $\cos\theta_1$.

We now compare (22) and (23) with the expression obtained by Kogelnik [16] for a transmitting phase hologram under the exact fulfillment of the Bragg condition. It is known that when $n_1$ is small, Kogelnik's formulas are also valid for small *d* [20]. Under these conditions, from Equation (42) in [16], we obtain the following expression for the amplitude of the diffracted wave.

$$-i\left(\frac{\cos\theta}{\cos\theta_1}\right)^{1/2}\sin\left[\frac{\pi n_1 d}{\lambda(\cos\theta\cos\theta_1)^{1/2}}\right]. \tag{24}$$

For a small *d*, by replacing the sine function in (24) with its argument, we obtain (23) up to a phase factor.

Let us check for expressions (22) and (23) the validity of the electrodynamic reciprocity theorem (Section 2.5 [15]), which, as applied to phase gratings, requires that when the wave vectors of the incident and diffracted waves are interchanged, and the sign of the grating vector is simultaneously changed, the diffraction efficiency does not change [12]. The diffraction efficiency η is given by [16]

$$\eta = \frac{|\cos\theta_1|}{\cos\theta}E_1(d)E_1^*(d). \tag{25}$$

For expressions (22) and (23) we obtain, respectively:

$$^{RN}\eta = \frac{k_0^2 n_1^2 d^2}{4\cos^3\theta}, \tag{26}$$

$$^{PT}\eta = \frac{k_0^2 n_1^2 d^2}{4\cos\theta\cos\theta_1}. \tag{27}$$

Obviously, expression (27) will not change when the angles θ and $θ_1$ are interchanged, whereas the value of expression (26) will change. Thus, the expression obtained using perturbation theory satisfies the reciprocity theorem[2], whereas the expression obtained based on the Raman-Nath expression does not. It is easy to see that, for expression (42) in [16], the reciprocity theorem is satisfied.

## 5. Discussion

A comparison of the expressions for the amplitude of the first diffraction order, obtained in the first approximation of the perturbation theory, in the two-wave approximation of the coupled-wave theory, and in the Raman-Nath theory showed that for a small thickness of the hologram, the first two expressions coincide with each other and differ from the third one. In addition, the reciprocity theorem is satisfied for the first two approaches but not for the third. We emphasize that the aforementioned difference in expressions is preserved for $d\to 0$ and $n_1 \to 0$. This difference becomes insignificant only when $\Lambda/\lambda$ increases, that is, when the condition of applicability of the Raman-Nath expression is satisfied. For the applicability of perturbation theory, this condition is not required, but in this case the amplitudes of the diffracted waves must be significantly lower than the amplitude of the incident wave. This allowed us to conclude that considering a thin hologram as a slide can lead to noticeable errors if the cosines of the incident and diffracted waves differ significantly.

---

[2] It is easy to show that the original expression (21) also satisfies the reciprocity theorem [12].

This conclusion is important not only for thin holograms but also for volume holograms when using the thin-grating decomposition method. It turns out that when using Expression (21) in the thin-grating decomposition method for the amplitudes of waves diffracted by each layer [2], the values of the amplitudes of the diffraction orders will be overestimated or underestimated, depending on the ratio of the cosines of the corresponding angles, which in both cases will be noticeable by the imbalance of energy. At the same time, when using Expression (23) in the calculations, the energy conservation law is fulfilled quite accurately [8].

Let us point to another interesting consequence of this consideration. Let the object field consist of several plane waves with amplitudes of $^j E$ and wave vectors $k_j$ lying in the $xz$-plane with different angles $\theta_j$ along the $z$-axis. Then, with linear registration and ideal reconstruction, the components of the reconstructed field are proportional to $\frac{1}{\cos\theta_j} {}^j E$ according to Expression (23). In other words, the object field restored in the first diffraction order is slightly distorted compared to the original one. This deduction can be easily generalized to the case of a continuous spatial-frequency spectrum.

## 6. Conclusion

The simplest consideration of the passage of a plane light wave through a thin hologram, like the passage of a light beam through a slide, leads to expressions for the amplitudes of diffraction orders coinciding with the expressions obtained in the Raman-Nath theory. We compared these expressions with the results obtained using other approaches. We obtained a solution for the task of light diffraction on a hologram of finite thickness with sinusoidal modulation of the refractive index in the first approximation of the perturbation theory. A rigorous formulation of the boundary value problem was used, taking into account the transmitted and reflected diffraction orders. The resulting expression for the first transmitted diffraction order was compared with a similar expression in Raman-Nath theory for hologram thickness $d$ tending to zero. To achieve this, the expressions are expanded to a series in $d$. The terms of the expansion of the first order in $d$ differ from that in the case of using perturbation theory, the denominator is $\cos\theta_1$, and in the case of using the Raman-Nath theory, $\cos\theta$. The difference between the expressions is preserved at $d\to 0$ and $n_1\to 0$, and disappears only as $\Lambda/\lambda$ increases. It is shown that the expression obtained in the perturbation theory approximation, in contrast to the expression obtained from the Raman-Nath expression, first, for a small $d$ coincides with the expression obtained by Kogelnik, and second, satisfies the electrodynamic reciprocity theorem. The use of expressions obtained using perturbation theory in the thin-grating decomposition method makes it possible to avoid energy imbalance. It is shown that when reconstructing an object field containing many plane waves with different inclination angles, the ratio of their amplitudes is somewhat distorted compared to the original one. The main conclusion is that considering a thin hologram as a slide can lead to noticeable errors if the cosines of the incident and diffracted waves differ significantly.

## Data Availability statement

No data were generated or analyzed in the presented research.

**Funding**

This research did not receive any specific grant from funding agencies in the public, commercial, or not-for-profit sectors. This research was conducted within the framework of the state task of Kotel'nikov Institute of Radio Engineering and Electronics of the Russian Academy of Sciences.

**Conflicts of interest**

The author has declared that no competing interests exist.

**Acknowledgment**

The author thanks P. A. Smolovich and A. P. Orlov for their help with the text editing.

**References**

[1] R. J. Collier, C. B. Burckhardt, and L. H. Lin, Optical Holography, Academic, New York, 1971.

[2] R. Alferness, Analysis of optical propagation in thick holographic gratings, Appl. Phys. 7 (1975) 29-33.

[3] R. Magnusson and T. K. Gaylord, Diffraction efficiencies of thin phase gratings with arbitrary grating shape, J. Opt. Soc. Am. 68 (1978) 806-809.

[4] R. Magnusson and T. K. Gaylord, Solutions of the thin phase grating diffraction equation, Opt. Commun. 25 (1978) 129-132.

[5] M. G. Moharam, T. K. Gaylord, and R. Magnusson, Criteria for Raman-Nath regime diffraction by phase gratings Opt. Commun. 32 (1980) 19-23.

[6] W. H. Carter, On some diffraction efficiency equations for a thick grating or hologram, Opt. Commun. 103 (1993) 1-7.

[7] C.V. Raman and N.S.N. Nath, The diffraction of light by high frequency sound waves: parts 1– 5, Proc. Indian Acad. Sci. 2 (1935) 406-412, 413-420; 3(1936) 75-84, 119-125, 459-465.

[8] I. V. Kartashova and A. M. Smolovich, The programs of the 3-D hologram diffraction efficiency calculations, Proceeding of NIKFI (All-Union Scientific Research Institute of Motion Pictures and Photography) 117 (1984) 24-30 (in Russian).

[9] O. B. Serov, A. M. Smolovich, and G. A. Sobolev, Thin holograms recorded with oppositely directed beams, Sov. Tech. Phys.Lett. 4 (1978) 95-96.

[10] M. G. Moharam and T. K. Gaylord, Rigorous coupled-wave analysis of planar-grating diffraction, J. Opt. Soc. Am. 7 (1981) 811-818.

[11] S. M. Rytov, Diffraction of light by ultrasonic waves, Izv. Akad. Nauk. SSSR Ser. Fiz. 2 (1937) 223-259 (in Russian).

[12] O. V Konstantinov, M. M Panakhov, and Y. F. Romanov, Electrodynamic perturbation theory for light diffraction from 3-D phase gratings, Opt. Spectrosc. 46 (1979) 551-554.

[13] R. Dusséaux, C. Faure, J. Chandezon, and F. Molinet, New perturbation theory of diffraction gratings and its application to the study of ghosts, J. Opt. Soc. Am. A 12 (1995) 1271-1282.

[14] H. J. Schmidt, M. Imlau, and K. M. Voit, Explaining the success of Kogelnik's coupled-wave theory by means of perturbation analysis: discussion. J. Opt. Soc. Am. A 31(2014) 1158-1166.

[15] F. T. S. Yu, Introduction to diffraction, information processing, and holography, MIT Press, Cambridge: Mass., 1973.

[16] H. Kogelnik, Coupled wave theory for thick hologram gratings, Bell Syst. Tech. J. 48 (1969) 2909-2947.

[17] M. Born and E. Wolf, Principles of optics: electromagnetic theory of propagation, interference and diffraction of light, Elsevier, 2013.

[18] L. I. Mandelshtam, Fifth lecture on optics, in: S. M. Rytov (Ed.), Lectures on optics, the relativity theory and quantum mechanics, Nauka, Moscow, 1972 pp. 32-42 (in Russian).

[19] R. Courant and D. Hilbert, Methods of Mathematical Physics, V. 1, Wiley-VCH Verlag GmbH & Company KGaA., Weinheim, 2004.

[20] F. G. Kaspar Diffraction by thick, periodically stratified gratings with complex dielectric constant, J. Opt. Soc. Am. 63 (1973) 37-45.